# Effect of oxygen plasma etching on graphene studied with Raman spectroscopy and electronic transport

**Isaac Childres**[1,2], **Luis A. Jauregui**[2,3], **Jifa Tian**[1,2], **Yong P. Chen**[1,2,3]
[1] Department of Physics, Purdue University, West Lafayette, IN, 47907, USA
[2] Birck Nanotechnology Center, Purdue University, West Lafayette, IN, 47907, USA
[3] School of Electrical and Computer Engineering, Purdue University, West Lafayette, IN, 47907, USA
E-mail: yongchen@purdue.edu

**Abstract.** We report a study of graphene and graphene field effect devices after exposure to a series of short pulses of oxygen plasma. We present data from Raman spectroscopy, back-gated field-effect and magneto-transport measurements. The intensity ratio between Raman "D" and "G" peaks, $I_D/I_G$ (commonly used to characterize disorder in graphene) is observed to increase approximately *linearly* with the number ($N_e$) of plasma etching pulses initially, but then decreases at higher $N_e$. We also discuss implications of our data for extracting graphene crystalline domain sizes from $I_D/I_G$. At the highest $N_e$ measured, the "2D" peak is found to be nearly suppressed while the "D" peak is still prominent. Electronic transport measurements in plasma-etched graphene show an up-shifting of the Dirac point, indicating hole doping. We also characterize mobility, quantum Hall states, weak localization and various scattering lengths in a moderately etched sample. Our findings are valuable for understanding the effects of plasma etching on graphene and the physics of disordered graphene through artificially generated defects.

**1. Introduction**

Graphene has received much attention recently in the scientific community because of its distinct properties and potentials in nanoelectronic applications. Many reports have been made on graphene's very high electrical conductivity at room temperature [1, 2] and its potential use as next-generation transistors [3], nano-sensors [4], transparent electrodes [5], and many other applications.

Plasma etching is a common tool used to pattern graphene nanostructures, such as Hall bars [2] and nanoribbons [6]. In addition, plasma etching is used to study how graphene's properties are affected by etching-induced disorder [7]-[9]. Other techniques that have been used to create artificial defects in graphene include ozone exposure [10], high-temperature oxidation [11] and energetic irradiation by positive ions [12]-[17], protons [18] and electrons [19]-[22].

Previous reports have characterized oxygen plasma's effect on graphene's field effect conductivity [7, 8], Raman spectra, and surface morphology as measured by atomic force microscopy (AFM) [7]. However, many aspects regarding the disorder generated by plasma etching in graphene remain to be better understood. For example, the initial stage of the disorder generation as reflected in Raman spectra (the so-called "graphite to nanocrystalline graphite" stage [23]) has not been fully characterized because of the corresponding narrow window for the amount of oxygen plasma exposure. The magneto-transport and carrier localization in plasma etched graphene have also not been studied before, to the best of our knowledge.

Here we present a study on graphene exposed to a various amount of oxygen plasma etching, focusing on its properties as characterized by Raman, field-effect measurements and magneto-transport measurements (including quantum Hall effect and weak localization).

## 2. Experimental procedure

Our graphene samples are fabricated by micromechanical exfoliation [1] of highly ordered pyrolytic graphite (HOPG, "ZYA" grade, Momentive Performance Materials) onto a p-doped Si wafer with 300 nm of $SiO_2$. Single-layer graphene flakes, typically around 100 $\mu m^2$ in size, are identified using color contrast with an optical microscope [24] and then confirmed with Raman spectroscopy (using a 532 nm excitation laser) [25]. Graphene field-effect devices are fabricated using electron-beam lithography. The electrical contacts (5 nm-thick chromium and 35 nm-thick gold) are fabricated by electron-beam evaporation.

Our graphene devices are exposed cumulatively to short pulses (~ ½ seconds) of oxygen plasma in a microwave plasma system (Plasma-Preen II-382, figure 1) operating at 100 W. A constant flow of $O_2$ is pumped through the sample space, and the gas is excited by microwave (manually pulsed on and off). The microwave generates an ionized oxygen plasma, which generates etched holes (observable by AFM similar as previously reported [7]) in graphene. The microwave-excited plasma pulses are applied to the samples cumulatively, and field-effect and Raman measurements are performed as soon as possible (~5 min) in ambient atmosphere and temperature after each pulse. The magneto-transport data are taken using a $He^3$ superconducting magnet probe several days after plasma exposure.

## 3. Results and Analysis

Figure 2(a) shows representative Raman spectra for a single-layer graphene device (sample "1") taken after multiple cumulative exposures to oxygen plasma etching pulses. Of particular interest is the disorder-induced characteristic Raman "D" peak (~1350 $cm^{-1}$) (other disorder-related peaks, such as the "D'" (~1620 $cm^{-1}$) and "D+D'" (~2940 $cm^{-1}$) peaks are also observed) [23]. The "D" peak initially rises with increasing exposure. After 14 pulses, the "D" peak has ~4 times the amplitude of the graphene's "G" peak (~1580 $cm^{-1}$), but with additional exposure, the "D" peak begins to attenuate along with the "G" and "2D" (~2690 $cm^{-1}$) peaks (both of which begin to attenuate since the first exposure). After 23 pulses, the "D" peak (while still significant) reduces to ~2 times the amplitude of the "G" peak and the "2D" peak is almost completely suppressed.

Figure 2(b) shows the progression of the peak intensity ratios ($I_D/I_G$ and $I_{2D}/I_G$) as functions of the number ($N_e$) of plasma-etching pulses. The dependence of the ratio of the intensities of the "D" and "G" peaks, $I_D/I_G$, on $N_e$ shows 2 different behaviors in the regimes of "low" ($N_e < \sim 14$) and "high" ($N_e > \sim 14$) defect densities (referred to as "nanocrystalline graphite" and "mainly $sp^2$ amorphous carbon" phases respectively in Ref. 23). $I_D/I_G$ begins at $\sim 0$ before the plasma exposure (the small "D" peak is likely due to the device fabrication process), increases with increasing $N_e$ in the low-defect-density regime to $\sim 4$ after 14 plasma exposures, then decreases with further increasing $N_e$ in the high-defect-density regime to $\sim 1.9$ for $N_e = 25$. On the other hand, the ratio of the intensities of the "2D" and "G" peaks, $I_{2D}/I_G$, continuously decreases with increasing $N_e$ from $\sim 3$ for $N_e = 0$ down to $\sim 0.3$ for $N_e = 25$. In figure 2(b), the data in the low-defect-density regime are fitted to

$$\frac{I_D}{I_G} = C \times N_e \quad \text{(dashed line)}, \tag{1}$$

and those in the high-defect-density regime are fitted to

$$\frac{I_D}{I_G} = \frac{C}{N_e} \quad \text{(dot-dashed line)}. \tag{2}$$

The inset of figure 2(a) shows $\log(I_D/I_G)$ versus $\log(N_e)$. A line fit of the data in the low-defect-density regime gives a slope of $\sim 1.1$, confirming the approximate linear relationship between $I_D/I_G$ and $N_e$ in that regime, as found in equation (1).

Figure 2(c) shows the full width of the "2D," "G" and "D" peaks as functions of $N_e$. The peaks widen with increasing $N_e$, especially at higher exposures.

Figure 3 shows the measured field-effect resistance of another typical graphene device (sample "2"). The back-gate-dependent resistance is shown for a various amount of exposure up to a dosage where the charge-neutral "Dirac" point (CNP) is no longer measurable within the gate voltage range used in our experiment. The inset of figure 3 shows the CNP as a function of $N_e$. The CNP voltage shifts to the positive direction with increasing plasma exposure, starting at 18 V before exposure and increasing to $\sim 100$ V after 4 pulses. This $\sim 80$ V of positive shift in the gate voltage corresponds to an increase in carrier concentration ($n$) by $\sim 6 \times 10^{12}$ cm$^2$.

A magneto-transport study is performed for a graphene field-effect device (sample "3") exposed to 2 oxygen plasma pulses. Prior to magneto-transport measurements, sample "3" is characterized by field-effect conductivity measurements and Raman spectroscopy before and after exposure, and the results are shown in figures 4(a) and 4(b) respectively. The CNP shifts from 2 V before exposure to 18 V after exposure, with the minimum conductivity ($\sigma_{min}$, taken as the conductivity at the CNP) decreasing from $\sim 275$ µS to $\sim 100$ µS. We can extract the field-effect mobility ($\mu_{FET}$) by examining the slope of the field-effect curve, conductivity ($\sigma$) versus back gate voltage ($V_g$), where $V_g$ is sufficiently far away from the CNP and the curve is in the linear regime, using [26]

$$\mu_{FET} = \frac{t}{\varepsilon} \times \frac{d\sigma}{dV_g}, \quad (3)$$

where $t = 300$ nm is the thickness of the SiO$_2$ and $\varepsilon = 3.9 \times \varepsilon_0 = 3.45 \times 10^{-11}$ F/m is the permittivity of the SiO$_2$. This gives room-temperature $\mu_{FET} \approx 400$ cm$^2$/Vs after exposure, compared to $\mu_{FET} \approx 9800$ cm$^2$/Vs before exposure.

The Raman spectra measured from sample "3" show the emergence of the characteristic "D" peak after exposure, with $I_D/I_G \approx 3$.

Figure 4(c) shows the longitudinal resistance (4-terminal measurements), $R_{xx}$, and the Hall resistance, $R_{xy}$, versus $V_g$ measured at a magnetic field $B = 18$ T (perpendicular to the graphene) and temperature $T = 0.5$ K for sample "3". $R_{xy}$ plateaus at $\pm h/2e^2$ and $h/6e^2$, where $h$ is the Planck constant and $e$ is the electron charge. The plateaus can also be seen at filling factors $\nu = \pm 2, \pm 6$ and $\pm 10$ in the magnetic field sweep at 0.5 K with 0 V on the back gate, figure 4(d), where $\nu$ is defined by [1]

$$\nu = \frac{nh}{eB}. \quad (4)$$

The quantized Hall plateaus correspond to

$$R_{xy}^{-1} = \pm \frac{\nu e^2}{h} \text{ with } \nu = 4\left(N + \frac{1}{2}\right). \quad (5)$$

We can also calculate mobility from the Hall effect measured in figure 4(d) using [26]

$$\mu_{HALL} = \frac{R_H}{\rho_{xx}(B=0)} \text{ where } R_H = \frac{dR_{xy}}{dB} \text{ at low } B. \quad (6)$$

This yields a post-exposure Hall mobility of ~600 cm$^2$/Vs. We also note a pronounced peak in $R_{xx}(B=0)$. This peak is due to weak localization, which is typically suppressed in low-disorder single-layer graphene devices [27].

$R_{xx}(B)$ measurements in low $B$ are also taken from sample "3" at various temperatures ranging from 0.5 K to 60 K and $V_g = 10$ V, shown in figure 5(a). These data characterize weak localization in the sample. Weak localization arises from the constructive interference between time-reversed multiple-scattering trajectories of phase coherent carriers, leading to coherent back-scattering of carriers and increasing the electrical resistance. When a perpendicular magnetic field is introduced to break time-reversal symmetry or the temperature is raised to destroy phase coherence, this interference is suppressed. From the low magnetic field data, we can extract the phase coherence length, $L_\varphi$, as well as the intervalley and

intravalley scattering lengths, $L_i$ and $L_*$ respectively, using [28-30]

$$\Delta \sigma_{xx}(B) = \frac{e^2}{\pi h} \times \left[ F\left(\frac{B}{B_\varphi}\right) - F\left(\frac{B}{B_\varphi + 2B_i}\right) - 2F\left(\frac{B}{B_\varphi + B_i + B_*}\right) \right]$$

with $F(z) = \ln(z) + \Psi\left(\frac{1}{2} + \frac{1}{z}\right)$ and $B_{\varphi,i,*} = \left(\frac{h}{8\pi e}\right) L_{\varphi,i,*}^{-2}$, (7)

where $\Delta \sigma_{xx}(B) = [\sigma_{xx}(B) - \sigma_{xx}(B=0)] - [\sigma_{xx}(B,T_h) - \sigma_{xx}(B=0,T_h)]$ and $\sigma_{xx}(B) = \frac{l}{w} \times \frac{1}{R_{xx}(B)}$.

The $l$ and $w$ denote the length and width of the graphene device and $\sigma(B,T_h)$ represents the magneto-conductivity at sufficiently high temperature (approximated using data at $T = 60$ K) for the weak localization feature to disappear. From the plots of these characteristic lengths as a function of temperature, seen in figure 5(b), we note $L_i$ and $L_*$ are relatively $T$-insensitive, averaging ~14 nm and ~4 nm respectively. We also note $L_\varphi$ decreases with increasing $T$ from ~23 nm at 0.5 K to ~10 nm at 60 K.

## 4. Discussion

The Raman spectrum taken before exposure indicates single-layer graphene with $I_D/I_G > 2$ and a symmetric Lorentzian 2D peak [25]. The evolution of $I_D/I_G$ as a function of induced plasma disorder can be attributed to a gradual evolution from the sp$^2$-bonded carbon found in graphene into amorphous carbon with appreciable sp$^3$ bonding [23]. This evolution is classified into 2 main regimes – a so-called "graphite to nano-crystalline graphite" phase characterized by increasing $I_D/I_G$ with increasing disorder (or "low-defect-density regime"), and a so-called "nanocrystalline graphite to manly sp$^2$ amorphous carbon" phase characterized by decreasing $I_D/I_G$ with increasing disorder (or "high-defect-density" regime) [23].

In the low-defect-density regime, an empirical formula known as the Tuinstra-Koenig relation [23, 31, 32] has been developed to extract the crystalline domain size $L_d$ (which also characterizes the average separation between defects):

$$\frac{I_D}{I_G} = \frac{C(\lambda)}{L_d}, \text{ where } C(\lambda) = (2.4 \times 10^{-10} \text{nm}^{-3}) \times \lambda^4 \qquad (8)$$

and $\lambda$ is the Raman excitation wavelength (532 nm in our case). In the high-defect-density regime, it has been proposed that $I_D/I_G$ versus $L_d$ can be fitted to the equation [23, 32]

$$\frac{I_D}{I_G} = D(\lambda) \times L_d^2 \qquad (9)$$

where the constant $D(\lambda)$ is obtained by imposing continuity between the two regimes.

A recent work [13], however, has suggested a new relationship between $I_D/I_G$ and $L_d$, in the low-defect-density regime as

$$\frac{I_D}{I_G} = \frac{C'(\lambda)}{L_d^2}. \qquad (10)$$

In our experiment, we assume the total exposure time to be proportional to the defect concentration, $1/L_d^2$, therefore

$$L_d \propto 1/\sqrt{N_e}.$$

We find that our data agrees with equation (9) in the high-defect-density regime (dot-dashed line fit in figure 2(b)). However, in the low-defect-density regime our data is better fitted to equation (10) than to equation (8). $C'(\lambda)$ is given to be 102 nm$^2$ for $\lambda = 514$ nm [13]. Assuming comparable $C'(\lambda)$ for our slightly different $\lambda$ (532 nm) in equation (10), we estimate $L_d \approx 5$ nm at the peak of $I_D/I_G$ (~4), a value similar to other reported values [13, 17].

The gradual decrease of the "2D" peak is also consistent with previous work [7, 9, 11]. The decreasing $I_{2D}/I_G$ versus $N_e$ is likely due to a combination of the positive doping of the graphene and the defect-induced suppression of the lattice vibration mode corresponding to the 2D peak.

The field-effect measurements show a strong positive shift of the CNP after plasma etching, most likely caused by p-doping molecules, e.g. water, attaching to the edges of etched holes [33]. If we use data in figure 3 as a guideline, 4 plasma pulses (which caused the CNP to up-shift by ~80 V, indicating a carrier density increase of ~6×10$^{12}$ cm$^2$) gives $I_D/I_G \approx 1.4$, from which we extract (using equation (10)) $L_d \approx 9.1$ nm, which corresponds to a defect concentration of ~8.4×10$^{11}$ cm$^2$. We therefore estimate an average of ~7 holes doped per defect. Such hole doping could also cause a small blue shift in the Raman "G" peak position [34], which, however, is not resolved within the resolution of our Raman measurements.

The electrical transport measurements show the oxygen plasma exposure significantly decreases the sample's $\sigma_{min}$ (at Dirac point) and mobility. Interestingly, even with this high level of disorder and low mobility (accompanied by pronounced weak localization), we still observe the half-integer quantum Hall effect in the form of well-developed plateaus in $R_{xy}$ corresponding to Landau level filling factors at 2, 6 and 10. We also note in figure 5(d) that although the quantum Hall plateaus in $R_{xy}$ are reasonably well-developed for $\nu = \pm 6$ and $\pm 10$, the corresponding $R_{xx}$ (which normally approaches 0 for quantum Hall effect states) still take substantial values (> ~0.8 $h/e^2$). More studies are needed to better understand this behavior and whether it could be related to, for example a proposed "dissipative" quantum Hall effect [35] or charge inhomogeneity (puddles) [33] in graphene.

Comparing the measured Raman spectrum from sample "3" after exposure with the spectra in figure 2 indicates the sample is in the low-defect-density regime. For $I_D/I_G \approx 3$ (figure 4(b)) we can calculate $L_d \approx 6$ nm from equation (10). This is on the similar order of magnitude as $L_i$ (~14 nm) and $L_*$ (~4 nm) extracted from weak localization fitting. Though it is still difficult to pinpoint exactly what kinds of disorder in graphene give rise to these scattering lengths, it has been proposed that atomically sharp defects largely contribute to $L_i$ while larger-length-scale disorder such as charged impurities can contribute to $L_*$ [10, 28].

## 5. Summary

In summary, we have studied the disorder in graphene caused by oxygen plasma etching. Raman spectra show a characteristic progression of $I_D/I_G$ with an increasing level of disorder, indicating an evolution from a graphene lattice to a more amorphous carbon phase in 2 distinct regimes, in which we can empirically extract defect length scales from $I_D/I_G$. The field effect in plasma-etched graphene shows a decrease in mobility and minimum conductivity, as well as a positive shift in the charge-neutral "Dirac" point voltage, indicating p-dopants attaching to the defect sites. Weak localization analysis reveals information about disorder-induced scattering, while well-developed quantum Hall plateaus are still observed in moderately etched samples. These results are valuable in understanding the effect of oxygen plasma etching on graphene as well as the nature of disorder in graphene.

**Acknowledgments**

This work has been partially supported by the National Science Foundation (ECCS#0833689), Department of Homeland Security (#2009-DN-077-15 ARI036-02) and by the Defense Threat Reduction Agency (HDTRA1-09-1-0047). Acknowledgment is also made to the donors of the American Chemical Society Petroleum Research Fund for partial support of this research. We also thank Thermo Scientific for assistance in Raman measurements. A portion of this work was carried out at the National High Magnetic Field Laboratory, which is supported by NSF Cooperative Agreement No. DMR-0084173, by the State of Florida and DOE. We thank Jun-Hyun Park and Eric Palm for experimental assistance. Yong P. Chen also acknowledges support from the Miller Family Endowment and IBM.

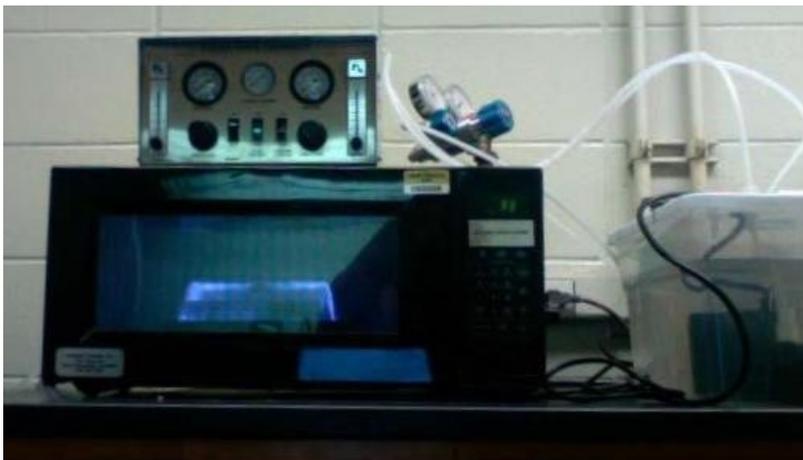

**Figure 1.** (color) Plasma-Preen II-382 system operating at 100W. Samples are placed under a Bell jar inside the modified microwave, where a controlled flow of oxygen is pumped through the jar and excited by the microwave radiation.

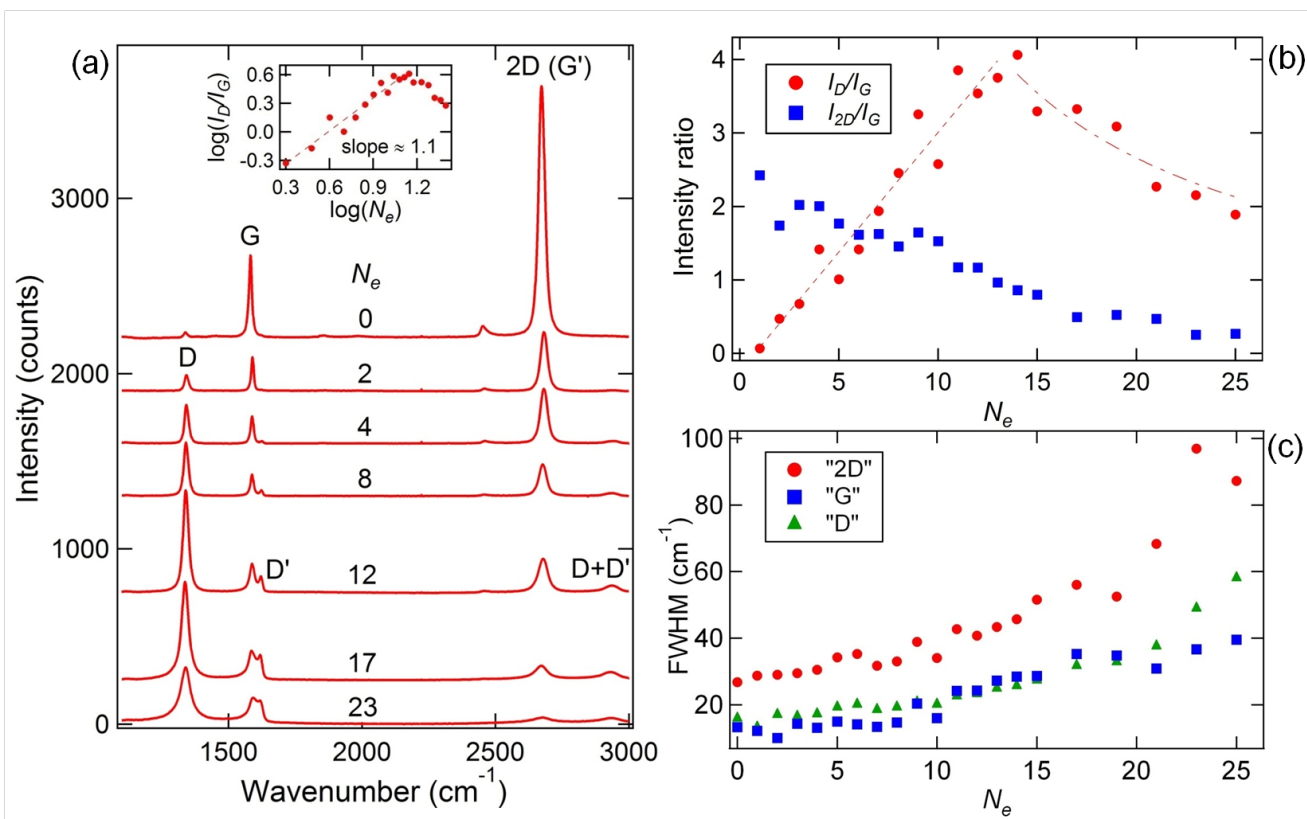

**Figure 2.** (a) Raman spectra of single layer graphene sample "1" after various numbers of accumulated oxygen plasma pulses, $N_e$. The spectra are offset vertically for clarity. (b) Ratios of Raman peak intensities, $I_D/I_G$ and $I_{2D}/I_G$ plotted against $N_e$. The dashed line is a fitting for $I_D/I_G = C \times L_e^{-2}$ (low-defect-density regime), and the dot-dashed line is a fitting for $I_D/I_G = C \times L_e^2$ (high-defect-density regime), where $L_e \propto 1/\sqrt{N_e}$. The inset of (a) shows $\log(I_D/I_G)$ against $\log(N_e)$. (c) The full width at half maximum of the "2D," "G" and "D" peaks plotted as functions of $N_e$.

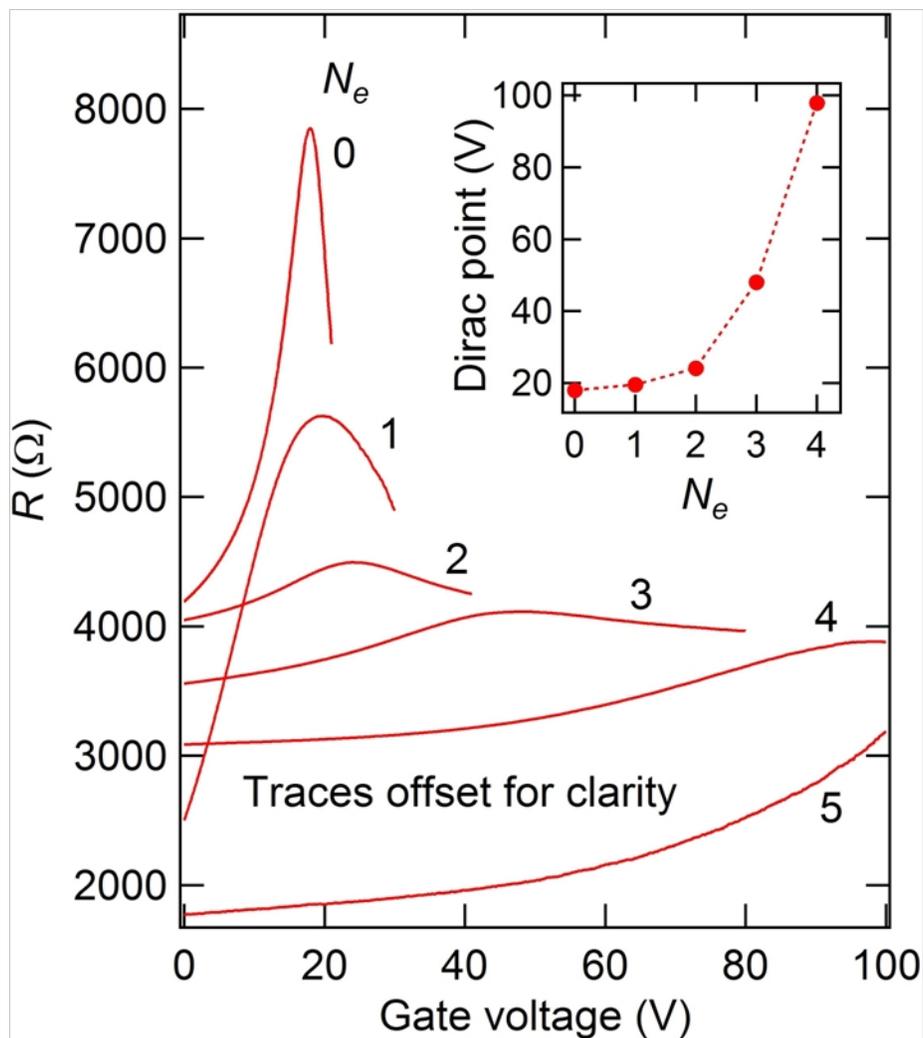

**Figure 3.** Measured resistance as a function of back gate voltage of single-layer graphene sample "2" after various numbers of accumulated oxygen plasma pulses, $N_e$. The measurement with $N_e = 0$ is 4-terminal, the measurement with $N_e = 1$ is 2-terminal and all others are 3-terminal. The traces are offset vertically for clarity. The inset shows the charge-neutral "Dirac" point as a function of $N_e$.

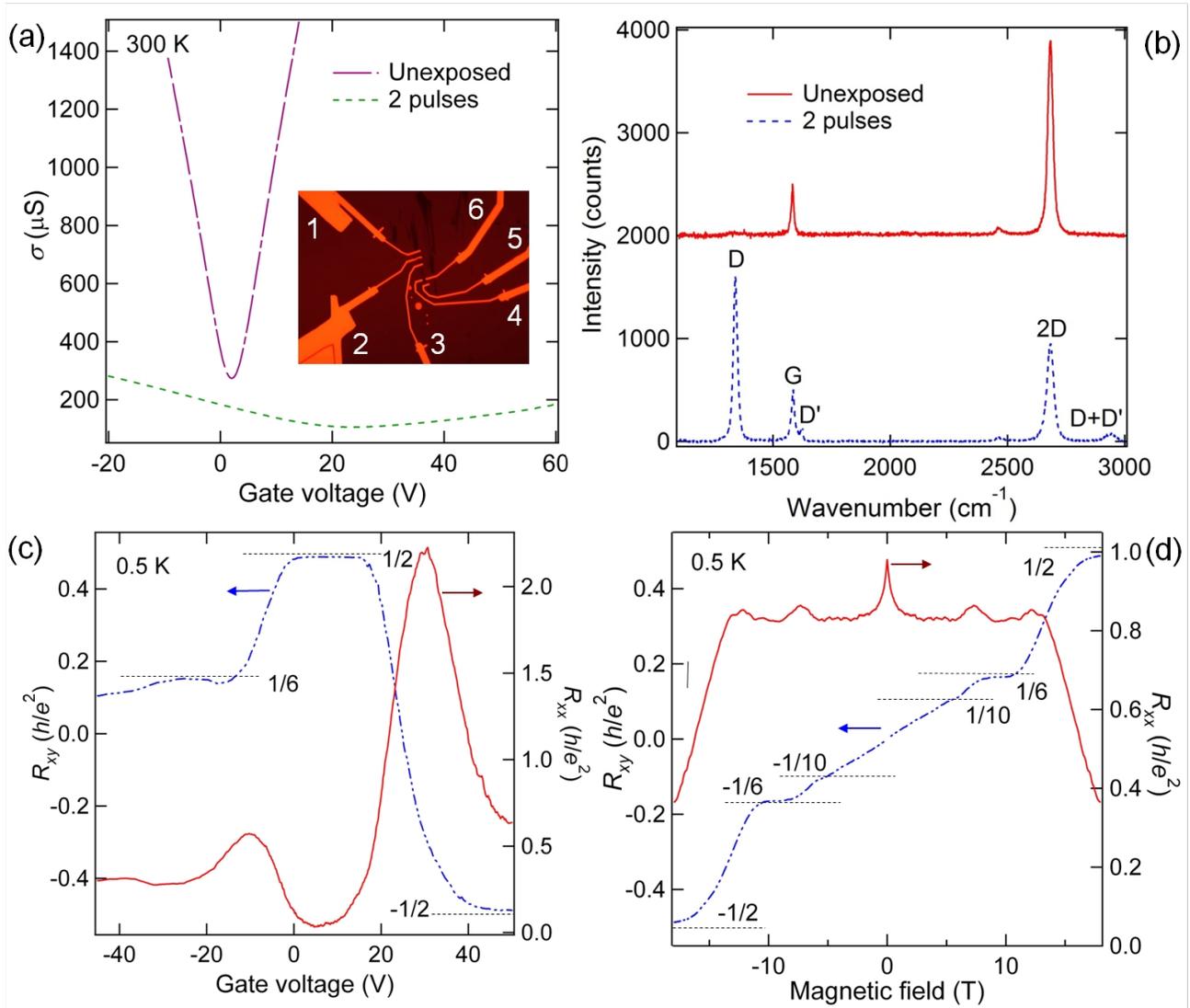

**Figure 4.** (a) Conductivity as a function of back gate voltage measured in single-layer graphene sample "3" before and after 2 oxygen plasma pulses. The inset of (a) is an optical image of sample "3". For all electrical measurements, the current was supplied from lead 1 to lead 5, with $R_{xx}$ measured from 3 to 4 and $R_{xy}$ from 6 to 4. (b) Raman spectra of sample "3" taken before and after plasma exposure. The laser excitation wavelength is $\lambda = 532$ nm. The spectra are offset vertically for clarity. (c) Measured resistance as a function of back gate voltage of post-exposure sample "3" at 18 T and 0.5 K. (d) Measured resistance as a function of magnetic field with no applied gate voltage under otherwise similar conditions as in (c). The dashed lines in (c) and (d) indicate the locations of the expected quantum Hall plateaus in $R_{xy}$ for filling factors 2, 6 and 10.

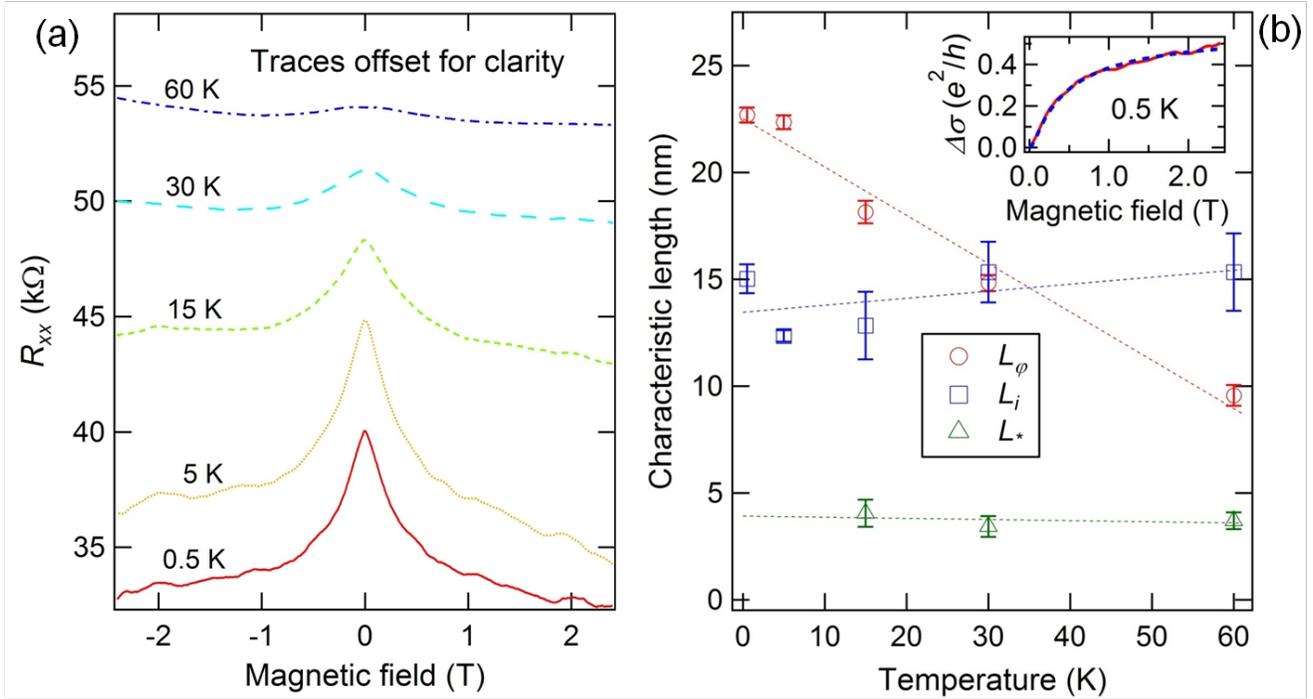

**Figure 5.** Temperature-dependent magneto-resistance showing weak localization in sample "3" after 2 pulses of oxygen plasma. (a) Magnetoresistance, $R_{xx}(B)$ measured at various temperatures. The traces are offset vertically for clarity. (b) Extracted characteristic lengths from weak localization fittings as functions of temperature. The inset shows magnetic-field-dependent $\Delta\sigma$ (solid line) and the result of fitting (dashed line) using equation (4) to extract $L_\varphi$, $L_i$ and $L_*$. $\Delta\sigma$ has been symmetrized between two opposite magnetic field directions.